\documentclass[12pt, amssymb,amsmath,aps,showpacs,floatfix,nofootinbib,showpacs,balancelastpage]{revtex4}
\usepackage{hangcaption,fancybox,feynmp}
\usepackage{indentfirst}
\usepackage{graphicx}
\usepackage{epsfig,subfigure,psfrag}
\usepackage{mathrsfs}
\usepackage{color}

\begin{document}

\include{newcmd}
%%%%%%%%%%%%%%%%%%%%%%%%%%%%%%%%%%%

\title{ Forward model with space-variant of source size  for reconstruction on x-ray radiographic image }
\author{Jin Liu$^{1}$\footnote{E-mail:ljin$\_$ifp@caep.ac.cn},Jun Liu$^{1}$, Yue-feng Jing$^{1}$,
Bo Xiao$^{1}$,Cai-hua Wei$^{1}$,Yong-hong Guan$^{1}$, Xuan Zhang$^{1}$}

\affiliation{ $ ^1$ Institute of Fluid Physics, CAEP, P. O. Box 919-105, Mianyang 621900, China }

\date{\today}
\maketitle

\begin{center}
{\bf Abstract}

\begin{minipage}{15cm}
{ \small
\hskip 0.25cm

 Forward imaging technique is the base of combined method on density reconstruction with the forward calculation and inverse problem solution. In the paper, we introduced the projection equation for the radiographic system with areal source blur and detector blur, gained the projecting matrix from any point source to any detector pixel with x-ray trace technique, proposed the ideal on gridding the areal source as many point sources with different weights, and used the blurring window as the effect of the detector blur. We used the forward projection equation to gain the same deviation information about the object edge as the experimental image. Our forward projection equation is combined with Constrained Conjugate Gradient method to form a new method for density reconstruction, XTRACE-CCG. The new method worked on the simulated image of French Test Object and experimental image. The same results have been concluded the affecting range of the blur is decreased and can be controlled to one or two pixels. The method is also suitable for reconstruction of density-variant object. The capability of our method to handle blur effect is useful for  all radiographic systems with larger source size comparing to pixel size. }

\end{minipage}
\end{center}

%\renewcommand{\baselinestretch}{1.2}
%\fontsize{12pt}{12pt}\selectfont

\newpage

%%  \linenumbers

%%\section{}
%%\label{}

\section{Introduction}\label{sec.introduction}

After the Comprehensive Test Ban Treaty (CTBT) \cite{Cunningham:2003}, the high-energy x-ray flash radiography has become an important technique to determine the edges and
density distributions of materials for the hydro-test. The intensity of the X-ray beam obeys Beer-Lambert law when it passes through an object,
and the Beer-Lambert law is the basic for density reconstruction\cite{Morris2003}. In the x-ray flash radiographic system with the goal to achieve 1\% density uncertainty, the 100ns, 2 kA, 20 MeV electron beam is used as the driven source to produce bremsstrahlung x-ray beam with average energy about 3-5 MeV, which has the most powerful penetrating capability. Although, the mean free path of French Test Object(FTO) exceeds 8.0\cite{Cunningham:2003}, the intensity of the  most powerful x-ray reduced to below one thousandth, and reduce the ratio of signal to noise. Besides the noise, the blur is an important factor for density reconstruction. The systematic blur of x-ray radiography consists of source size and detector blur. The space charge effect of the kA level current limits the reduction of the source size, and the full width of half maximum (FWHM) of source size is over 1.0 mm for the leading machines such as DARHT and Dragon-I\cite{Davis2006,Ding:2005,Jiang2012}. The detector blur of 20 mm CsI plate is about 2.1 mm.

Some experimental measurement is carried out to reduce the impact of the blur on the image quantity. We can optimize the system magnification to make the system blur as small as possible. And the optimized magnification \textit{m} is satisfied that\cite{Burns:1999,LJIN2012}
\begin{equation}\label{eq.magification}
    m =1+\frac{\textrm{FWHM}^{2}_{\textrm{D}}}{\textrm{FWHM}^{2}_{\textrm{S}}},
\end{equation}
After that, the system blur is still over 1.0 mm, much more lager than one pixel size at object plane. In order to gain better result, the blur factor must be considered in density reconstruction.

At present, the blur effect of the areal source is usually handled as a space-invariant blur as the detector blur in the most popular reconstruction methods includes analytical methods (such as Abel transform, filter back projection(FBP)) , iterative methods(such as ART). In hydro-test radiography, the statistic method Bayesian Inference Engine is used at Los Alamos National Laboratory(LANL), which is a type of constrained method for the statistic term of the object information equals a constraint. Meanwhile, the LANL forward model for density reconstruction was proposed\cite{Aufderheide2002}. In the model, the source size effect is also considered space-invariant as the detector blur. They payed more attention on the attenuation coefficient of spectrum effect rather than the systematic blur.

In this paper, we proposed our forward reconstruction model with space-variant of source size. The space-variant of the source size is considered, and it is achieved using x-ray trace technique and gridded source methods. Our model was verified by comparing between experimental image and calculated image.

\section{theory} \label{sec.theory}

\subsection{forward model}

	In ref\cite{Aufderheide2002}, the LANL forward model is
\begin{equation}\label{eq.LANL_model}
    y_{i}=\sum_{j} \textbf{B}_{ij}\textbf{B}'_{i,j}[S_{j}\exp (-\sum_{l}\sum_{k}\mu_{ljk}a_{ik}x_{k})+\zeta_{j}].
\end{equation}
Where $\textbf{B}_{ij}$, $\textbf{B}'_{i,j}$ are the elements of  matrix of detector blur and source blur, respectively. The $S_{j}$ is the photon number of the jth energy bin. The $u_{ijk}$ is the mass absorption of material $l$ in voxel $k$ at energy bin $j$. $a_{ik}$ is ikth element of the projection matrix $\textbf{A}$, $\zeta_{j}$  is scatter radiation at detector i.
In Equ.\ref{eq.LANL_model}, the source blur is a convolution matrix at detector plane. It means that the source blur is space-invariant.
In fact, the projection matrix $\textbf{A}$ is different via the change of the source position. So we proposed our forward model(IFP model)with many weighted source points. The model is

\begin{equation}\label{eq.proposed_model}
    \textbf{X}_{T}=\textbf{B}_{det}\sum_{i}[w_{i}X_{0}\exp(-\textbf{A}^{(i)}\textbf{x})]+X_{S}+\xi
\end{equation}
Where $\textbf{X}_{T}$ is total dose at detector plane, $\textbf{B}_{det}$ is detector blur matrix, $w_{i}$ is the ith source weight, $X_{0}$ is dose for empty field at detector plane, $\textbf{A}^{i}$ is the projection matrix of the ith source, $X_{S}^{i}$ is scatter dose and $\xi$ is noise.

If we consider that source blur is space-invariant, we can just use the blur matrix $\textbf{B}_{sys}$ to demonstrate system blur, which is equal to $\textbf{B}_{ij}\textbf{B}'_{i,j}$ of Equ.\ref{eq.LANL_model} as LANL model does. And we use $\textbf{B}_{sys}$ as system blur to instead of $\textbf{B}_{det}\sum_{i}w_{i}$ in Equ.\ref{eq.proposed_model}, we obtained the projection equation with space-invariant blur as LANL model

\begin{equation}\label{eq.proposed_model_space_invariant}
    \textbf{X}_{T}=\textbf{B}_{sys}[X_{0}\exp(-\textbf{A}^{(i)}\textbf{x})]+X_{S}+\xi
\end{equation}

\subsection{projection matrix}
\begin{figure}[htpb]
\centering
  \includegraphics[height=1.5in]{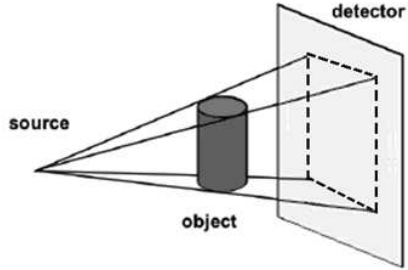}\\
  \caption{The layout of x-ray radiography system with cone beam} \label{fig.layout}
\end{figure}

The different source point \textit{i} induces different projection matrix $\textbf{A}^{(i)}$. The layout of x-ray radiography is shown in Fig.\ref{fig.layout}.

\begin{figure}[h]
\centering
 \subfigure[grids along rotational axis]{
   \label{fig.grid_along_rotational_axis} %% label for first subfigure
    \includegraphics[height=2.0in]{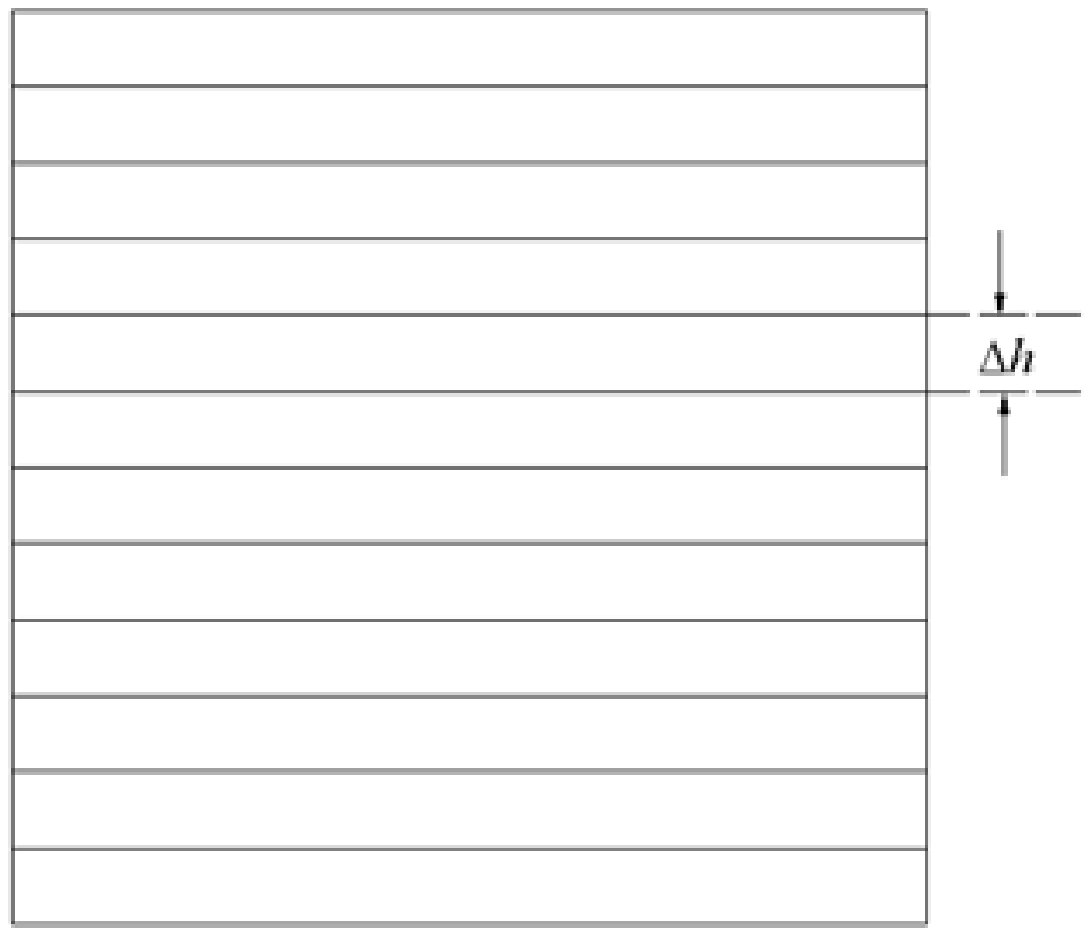}} \hspace{0.1cm}
     \subfigure[grids vertical to rotational axis]{
    \label{fig.grid_circles} %% label for second subfigure
    \includegraphics[height=2.0in]{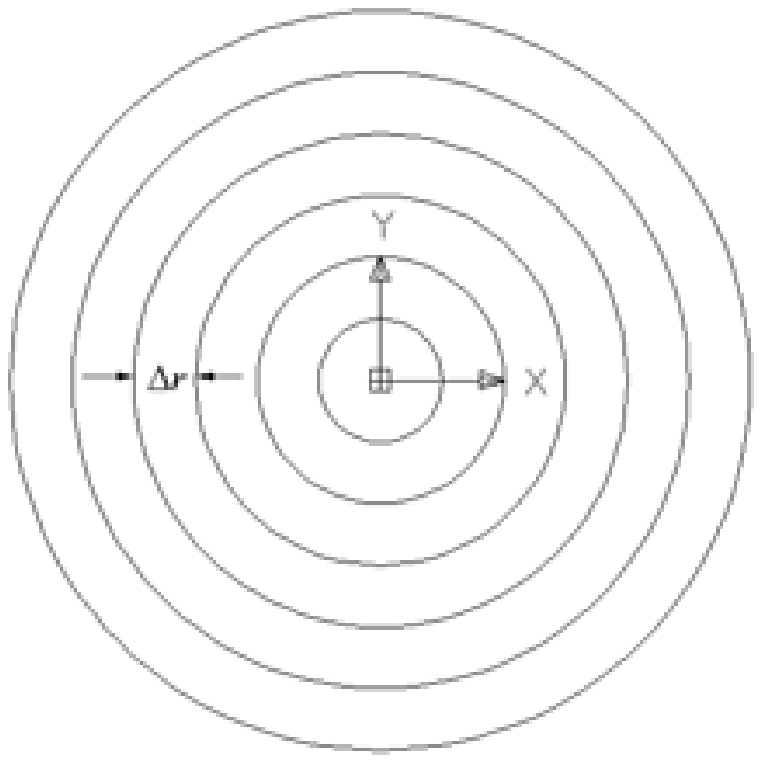}} \hspace{0.1cm}
\setlength{\abovecaptionskip}{2pt}
\setlength{\belowcaptionskip}{-6pt} \caption{The grids of rotational object}
\label{fig.gridded_object} %% label for entire figure
\end{figure}

The object of a single view radiographic system is rotational symmetrical and can be divided into 2D grids(see Fig.\ref{fig.gridded_object}). Along the rotational axis, the grids are parallel slabs with thickness $\Delta h$ , and the grids are concentric circles with radius increment $\Delta r$  on the Fig.\ref{fig.gridded_object}.

\begin{figure}[h]
\centering
 \subfigure[intersection with horizontal grids]{
   \label{fig.intersection_with_horizontal_grids} %% label for first subfigure
    \includegraphics[height=1.5in]{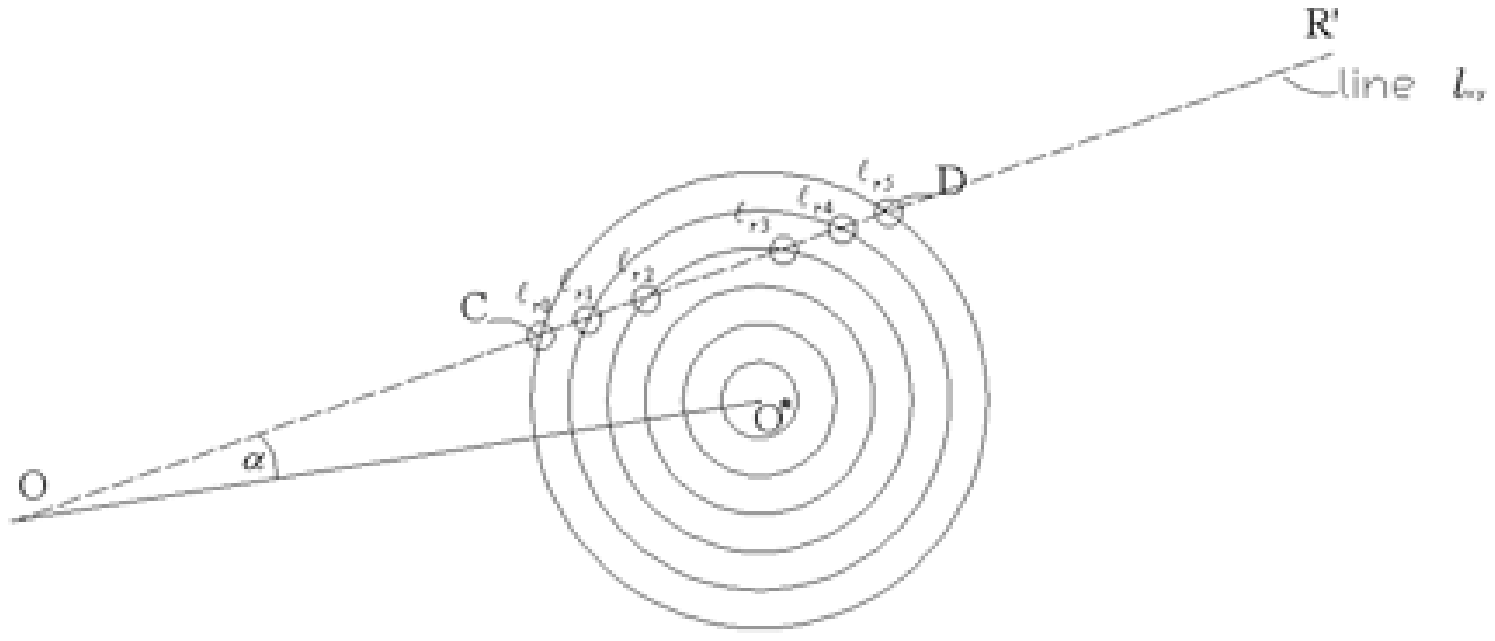}} \hspace{0.1cm}
     \subfigure[intersection with vertical grids]{
    \label{fig.intersection_with_vertical_grids} %% label for second subfigure
    \includegraphics[height=1.0in]{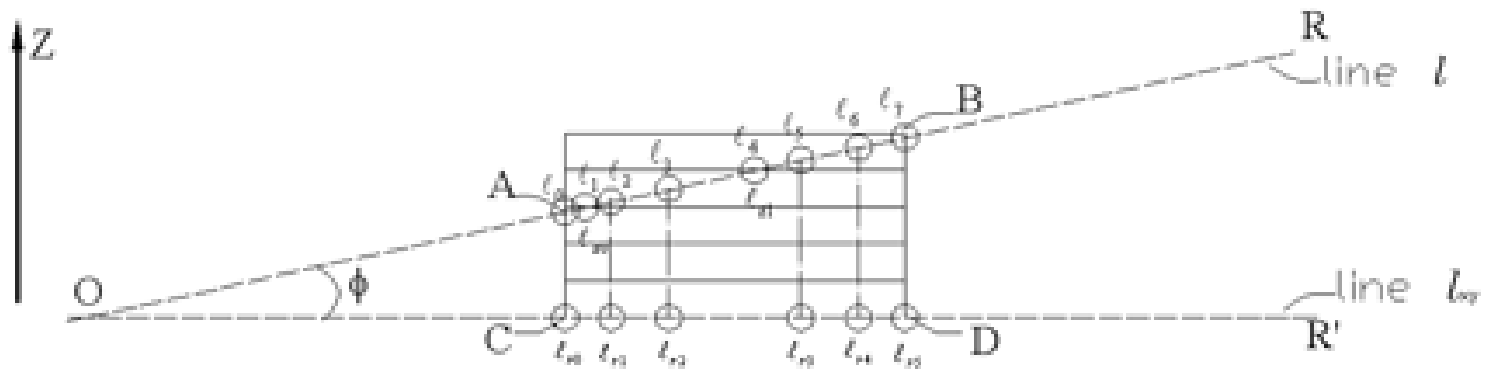}} \hspace{0.1cm}
\setlength{\abovecaptionskip}{2pt}
\setlength{\belowcaptionskip}{-6pt} \caption{The intersection of the x-ray and gridded object}
\label{fig.intersections} %% label for entire figure
\end{figure}

After the source position and object model was determined, We used x-ray trace method to calculate the matrix $\textbf{A}^{i}$. With arbitrary source point O ($x_{\textrm{O}},y_{\textrm{O}},z_{\textrm{O}}$), detected point R($x_{\textrm{R}}, y_{\textrm{R}},z_{\textrm{R}}$) and its projecting point R'($x_{\textrm{R}}, y_{\textrm{R}},z_{\textrm{O}}$) at vertical plane at $z=z_{\textrm{O}}$. The X-ray $\overrightarrow{\textrm{OR}}$ intersects with the object and its projection  is $\overrightarrow{\textrm{OR}'}$ in plane $z=z_{\textrm{O}}$. At this plane, X-ray  intersects with the concentric circles of the gridded object and forms \textit{n}+1 points of intersection. From the Fig.\ref{fig.intersection_with_horizontal_grids} the distance from the source point $\textrm{O}$ to the intersecting point is

\begin{equation}\label{eq.distance}
    \ell_{rp}=|\overrightarrow{\textrm{OO}^{*}}|\cos\alpha\pm\sqrt{r^{2}-(|\overrightarrow{\textrm{OO}^{*}}|\sin\alpha)^{2}}
\end{equation}
Where $\textrm{O}^{*}$ is the center point of the object, $|\overrightarrow{\textrm{OO}^{*}}|$  is the length of the vector $\overrightarrow{\textrm{OO}^{*}}$ , $\alpha$  is the angle between vectors $\overrightarrow{\textrm{OO}^{*}}$  and $\overrightarrow{\textrm{OR}'}$. Using Equ.\ref{eq.distance}, we obtained ordered distances of the $\ell_{rp}$'s and they can be described as $ \ell_{r_{0}},\ell_{r_{0}},\cdots,\ell_{r_{n}}$. Then  projecting the X-ray $\overrightarrow{\textrm{OR}'}$  back to   $\overrightarrow{\textrm{OR}}$,  the really lengths of the X-ray $\overrightarrow{\textrm{OR}}$  intersects of the object are gained as $ \ell_{r_{0}}/\cos\phi,\ell_{r_{0}}/\cos\phi,$ $\cdots,\ell_{r_{n}}/\cos\phi$ .Where $\cos\phi$ is the cosine of the angle between the  vectors $\overrightarrow{\textrm{OR}'}$ and  $\overrightarrow{\textrm{OR}}$ .

Then we calculated the distance from the source point $\textrm{O}$ to intersecting point of the x-ray and gridded object at  the surface of slabs on Fig.\ref{fig.intersection_with_vertical_grids}. Because the $q$th slab position is at plane $z=z_{\textrm{q}}$, the distance $\ell_{rq}$  is
\begin{equation}\label{eq.z_distance}
    \ell_{zq}=(z_{\textrm{q}}-z_{\textrm{O}})/u,  \texttt{if}  \ell_{r_{0}}/\cos\phi \leq \ell_{zq}\leq \ell_{r_{n}}/\cos\phi
\end{equation}
Where $u$ is the cosine of the angle between the vector $\overrightarrow{\textrm{OR}}$ and z-axis. Now, we put all the distance into a set $\{ \ell_{r_{0}}/\cos\phi,\ell_{r_{0}}/\cos\phi,\cdots,\ell_{r_{n}}/\cos\phi,\ell_{z0},\ell_{zq},\cdots, \ell_{zm}\}$ ,called C. After putting the all data in the set C in ascending sort, we got the right set D $ \{\ell_{0},\ell_{1},\cdots,\ell_{n+m+1}\}$ to gain projection elements $\textbf{A}^{(i)}_{kl}$ , and it follows
\begin{equation}\label{eq.matrix_element}
    \textbf{A}^{(i)}_{kl}=\ell_{n'+1}-\ell_{n'}
\end{equation}
Where the superscript $i$  of the projection matrix elements the index of the source points, the subscript $k$ is the index of the detector pixels and the subscript $l$ is the index of the object voxel.

\subsection{source position and weight}\label{subsec.source}
The forward model is the base for the combined reconstruction method including the forward projection and solution of inverse problem.
In the most popular method, the source is usually considered as an ideal point. And it is not suitable for an areal source.
So it is necessary to simulate or describe the source distribution in detail other than a point.

The best methods to simulate/calculate the radiographic image is Monte Carlo (MC)method. In which the source size is achieved by sampling source point according to the source size distribution function $f(r,\theta)$. The advantage of the MC method is that the description of the source is correct enough as the size distribution function, and the space- variant is considered naturally. But the speed for the simulating process is very slow so that the MC method is not suitable to implement into density reconstruction. Even so, the idea on description of the source position is great, and it can be expressed as weighted gridded source.In the Fig.1, the source is divided into several concentric circles with same radius increment $\Delta r$. The source point is determined with the angle increment  $\pi /2n$ and the source is given different weight according to source distribution function. The final source can be expressed as follows.

\begin{equation}\label{eq.source_parameters}
\left\{%
\begin{array}{ll}
     x_{i}=(j-0.5)\Delta r\cos\theta_{i},  \\
     y_{i}=(j-0.5)\Delta r\sin\theta_{i}, \\
     w_{i}=\int_{\theta_{i-0.5}}^{\theta_{i+0.5}}\int_{(j-1)\Delta r}^{j\Delta r}f(r,\theta)rdrd\theta
\end{array}%
\right.
\end{equation}
where j equals $\textrm{int}[(i-1)/4n]+1$, $x_{i}$, $y_{i}$ and $w_{i}$ are source x-axis, y-axis coordinations and weights, and $\theta_{i}$ is
\begin{equation*}
   \theta_{i}=[(i-1)\pi /2n],
\end{equation*}

\begin{figure}[htpb]
\centering
  \includegraphics[height=2.0in]{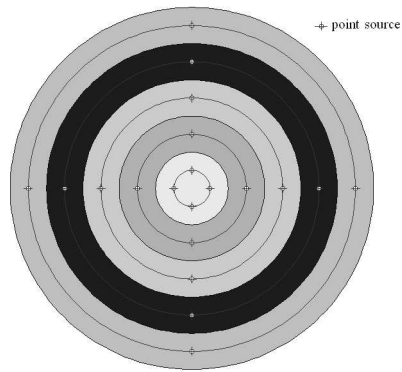}\\
  \caption{The diagram of the gridded areal source with angle increment  $\pi/2$} \label{fig.gridded_source}
\end{figure}

\subsection{matrix of detector blur}
The effect of the blur can be expressed as a window, just as a filter window. The windows size is N*N, and N is equal to the devision of 3sigma of the blur size to pixel size. The element of window matrix is
\begin{equation}\label{eq.matrix_element_detector_blur}
    b_{ij}=\int_{x_{i}}^{x_{i+1}}\int_{y_{j}}^{y_{j+1}}B(x,y)dxdy
\end{equation}
where $B(x,y)$ is the distribution of the detector blur.

\subsection{space-variant of source size effect}
The space-variant of the size effect presents in difference of the projection matrixes from different point source. We accounted the elements numbers of the projection matrix for x-ray radiography of an Object.The parameters of the x-ray radiographic system are that Full Width of Half Maximum of source is 1.5 mm, the Object locates 200.0 cm downstream from the source, the detector locates 300.0 cm and the voxel size of the object is 0.2 mm* 0.2 mm.  As described in section\ref{subsec.source}, we divided the areal source into 24 points(six circles and four points per circle). From the accounted result of projection matrixes in table\ref{table:matrix_element}, we found that the difference of the matrix elements number is larger than 3\%. So the space-variant of the size effect exits and it becomes obvious via the increment of the source size.

\begin{table}[!h]
\tabcolsep 0pt \caption{number of matrix elements} \vspace*{-12pt}
\begin{center}
\def\temptablewidth{0.85\textwidth}
{\rule{\temptablewidth}{1pt}}
\begin{tabular*}{\temptablewidth}{@{\extracolsep{\fill}}cccccccc}
\label{table:matrix_element}
& No.   & account & No.   & account & No.  & account \\
\hline
&      \#1 &  30745264 &      \#9 &  30260432  &      \#17 &  29778914 \\
\hline
&      \#2 &  33740127 &      \#10 &  32941718  &     \#18 &  32076213 \\
\hline
&      \#3 &  30745264 &      \#11 &  30260432  &     \#19 &  29778914 \\
\hline
&      \#4 &  33740127 &      \#12 &  32941718  &     \#20 &  32076213 \\
\hline
&      \#5 &  30502316 &      \#13 &  30019524  &     \#21 &  29539134 \\
\hline
&      \#6 &  33356949 &      \#14 &  32509435  &     \#22 &  31639597 \\
\hline
&      \#7 &  30502316 &      \#15 &  30019524  &     \#23 &  29539134 \\
\hline
&      \#8 &  33356949 &      \#16 &  32509435  &     \#24 &  31639597 \\
            \end{tabular*}
       {\rule{\temptablewidth}{1pt}}
       \end{center}
       \end{table}

\section{verification} \label{sec.verification}
\begin{figure}[h]
\centering
 \subfigure[]{
   \label{fig.balanced_exp_image} %% label for first subfigure
    \includegraphics[height=1.4in]{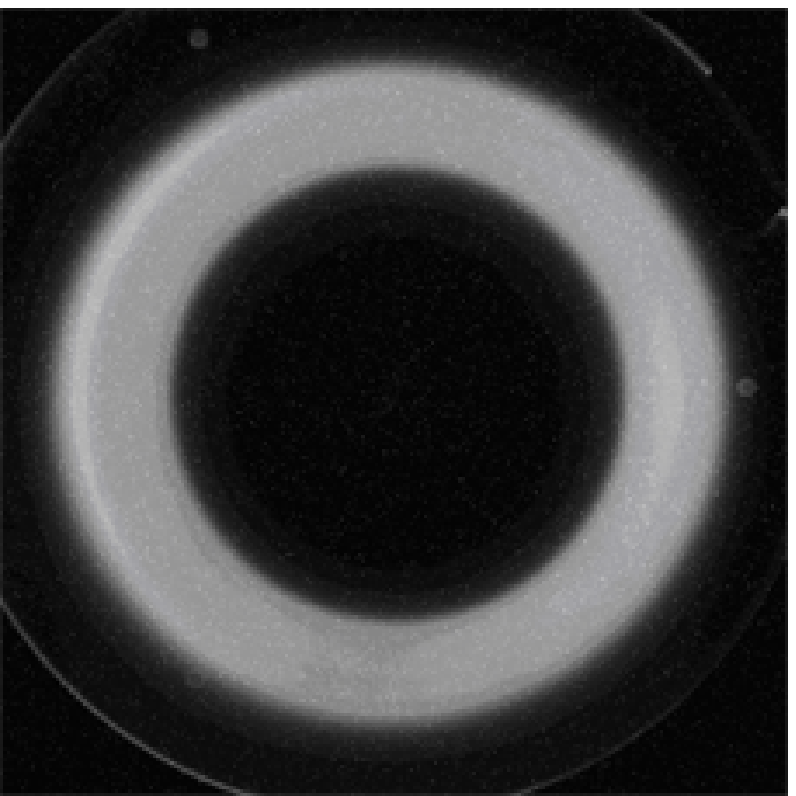}} \hspace{0.1cm}
     \subfigure[]{
    \label{fig.denoised_profile} %% label for second subfigure
    \includegraphics[height=1.4in]{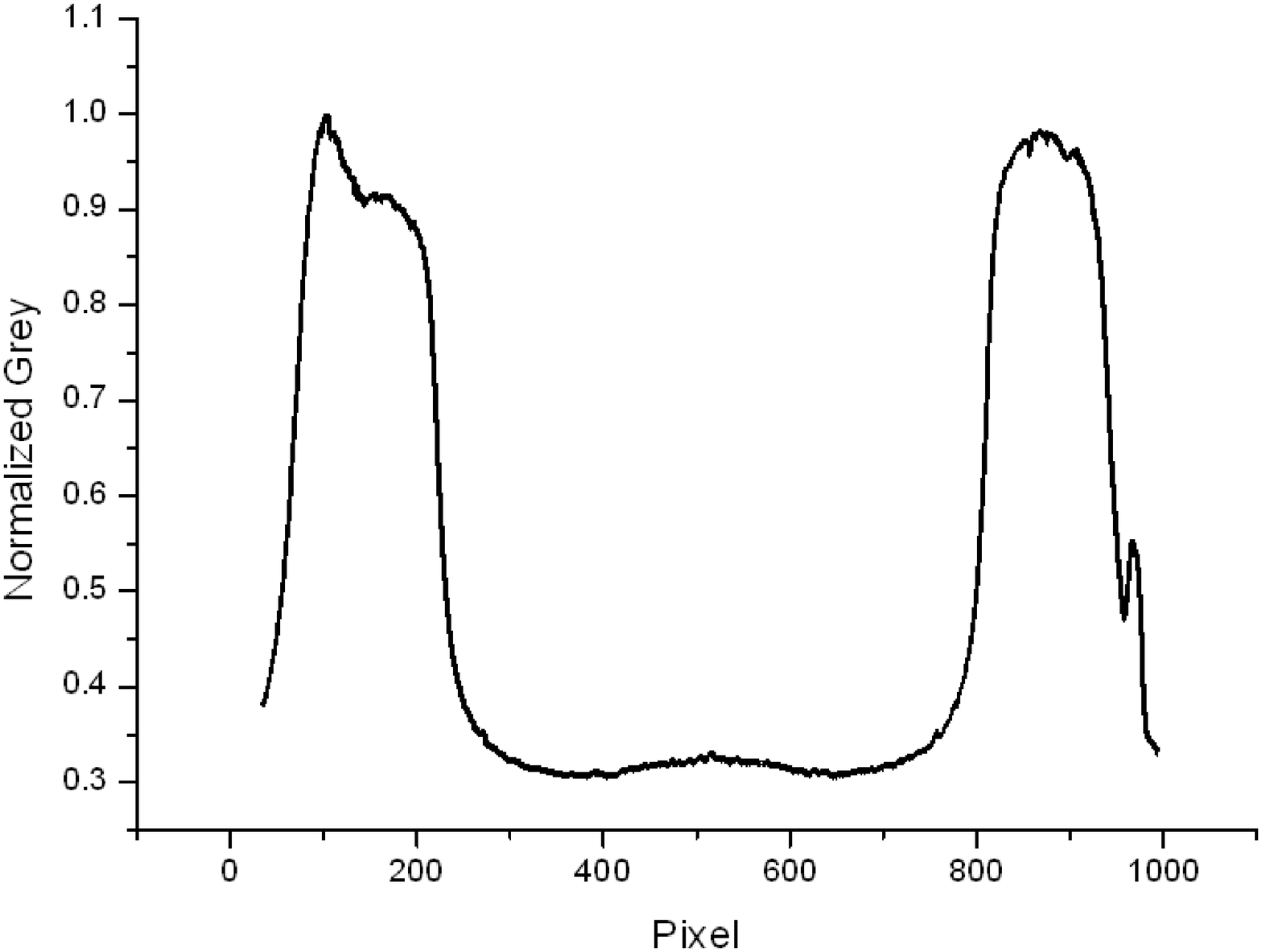}} \hspace{0.1cm}
     \subfigure[]{
    \label{fig.calculated_image} %% label for second subfigure
    \includegraphics[height=1.4in]{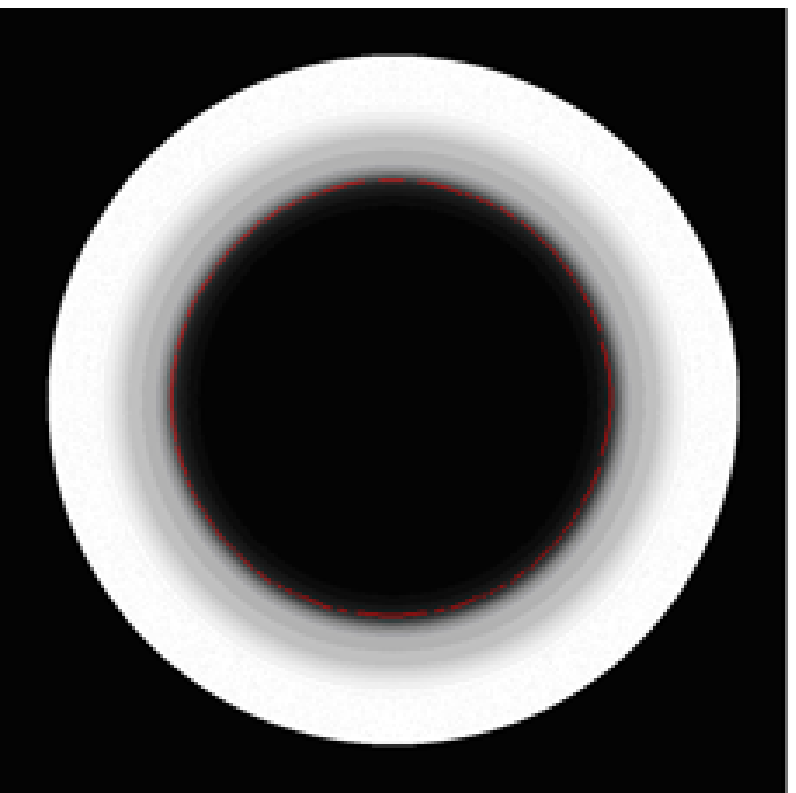}} \hspace{0.1cm}
\setlength{\abovecaptionskip}{2pt}
\setlength{\belowcaptionskip}{-6pt} \caption{The balanced experimental image(left), denoised profile(middle) and forward calculated image of a spherical object(right)}
\label{fig.intersections} %% label for entire figure
\end{figure}

	One method to prove the correction of our model is the consistence of the edge position between an experimental image and calculated image with our proposed forward model. The parameters used in calculating the simulated image can be measured with Roll-bar method, the source size(FWHM) is 1.5 mm , and detector blur is 2.1 mm. The experimental object is similar to FTO. The experimental image and its denoised central  profile were showed on Fig.\ref{fig.balanced_exp_image}-Fig.\ref{fig.denoised_profile}. The calculated image is Fig.\ref{fig.calculated_image}. After using the gradient method to obtain the outsider edge of the tungsten layers, we got the deviation of the detected edge from the given value is about 2.0 mm whether for the experiment image or calculated image. And the detected edge is shown in red line on Fig.\ref{fig.calculated_image}.

\section{application}\label{sec.application}

The forward model combined with reconstruction method, such as conjugate constrained gradient(CCG)\cite{LJIN2015}, to form a more precise method for density reconstruction of penetrating radiography and the combined method is called as XTRACE-CCG. The CCG method is stably convergent. The XTRACE-CCG is applied for density reconstruction of a simulated image and an experimental image.

\subsection{reconstruction of a simulated image}\label{subsec.rec0}

\begin{figure}[h]
\centering
     \subfigure[the total view]{
    \label{fig.total_view} %% label for second subfigure
    \includegraphics[height=1.8in]{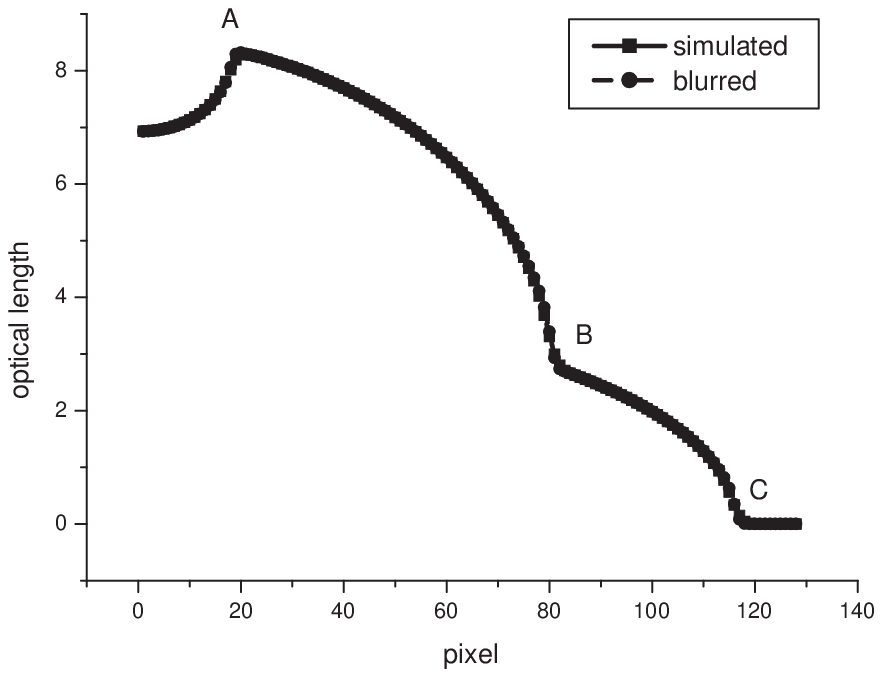}} \hspace{0.3cm}
     \subfigure[The magnified part of Fig.\ref{fig.total_view}]{
    \label{fig.magnified_part} %% label for second subfigure
    \includegraphics[height=1.8in]{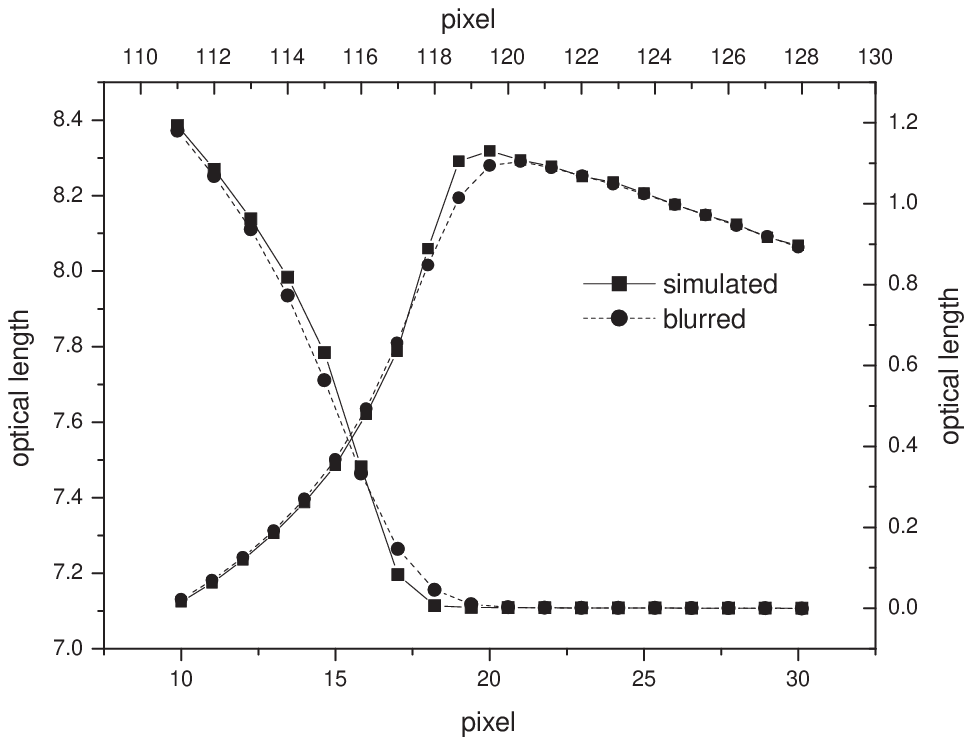}} \hspace{0.3cm}
\setlength{\abovecaptionskip}{2pt}
\setlength{\belowcaptionskip}{-6pt} \caption{The central line of the optical length with or without blur.
The solid line with square is FXRMC simulated result and the dash line with circle is FXRMC simulated result convolving with detector blur. }
\label{fig.optical_length} %% label for entire figure
\end{figure}

\begin{figure}[h]
\centering
     \subfigure[the total view]{
    \label{fig.total_view1} %% label for second subfigure
    \includegraphics[height=1.8in]{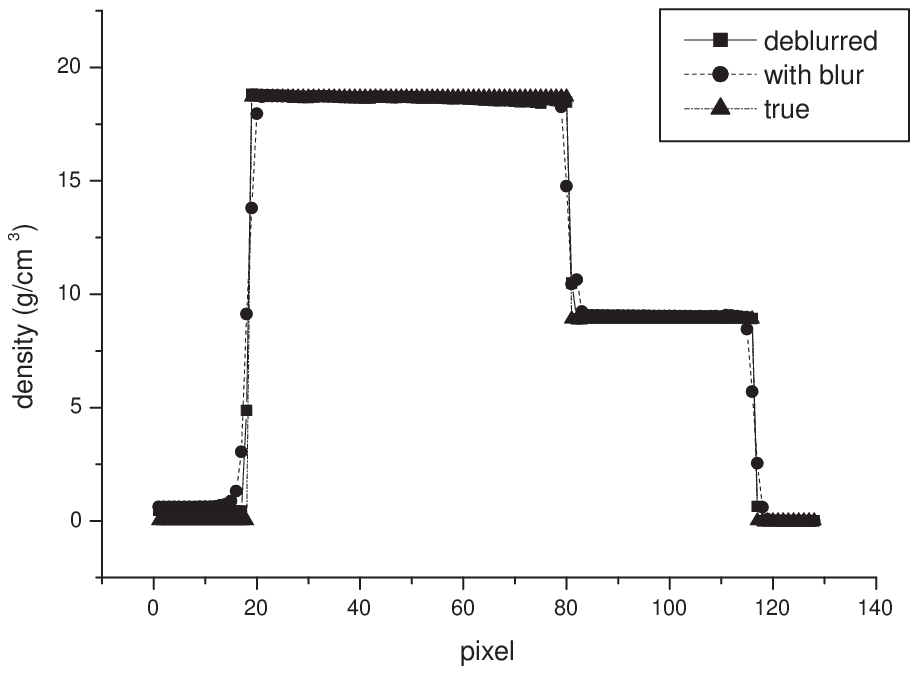}} \hspace{0.3cm}
     \subfigure[The magnified part of Fig.\ref{fig.total_view1}]{
    \label{fig.magnified_part1} %% label for second subfigure
    \includegraphics[height=1.8in]{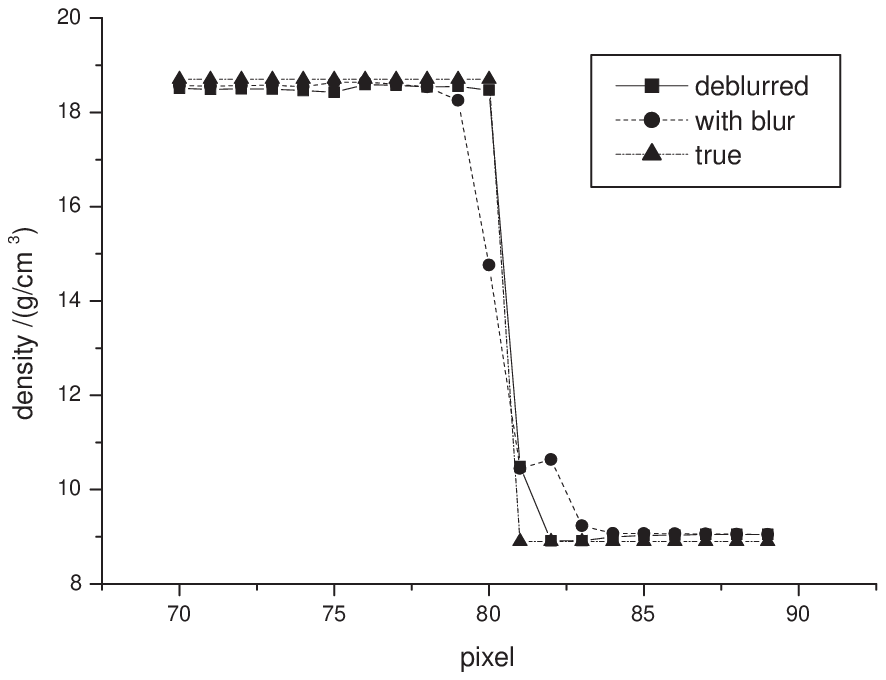}} \hspace{0.3cm}
\setlength{\abovecaptionskip}{2pt}
\setlength{\belowcaptionskip}{-6pt} \caption{The central line of reconstructed density from a simulated image with different forward equations. The solid line with square is deblurred
 result of our forward model, the dash line with circle is result without consideration of blur and the dash dot line with triangle is true value.}
\label{fig.rec_images} %% label for entire figure
\end{figure}

We used the XTRACE-CCG method to reconstruct the density from the simulated direct x-ray information. The parameters of simulated x-ray radiography layout are that the French Test Object locates 200.0 cm downstream from the source and the detector locates 300.0 cm. The simulated image is gained by that FXRMC(a simulated code for photon radiography developed by institute of fluid physics) simulates the direct x-ray information with areal source effect and the direct x-ray information convolved the Gaussian detector blur(FWHM is 2.1 mm). The central line of the simulated image is shown on Fig.\ref{fig.optical_length}. The Fig.\ref{fig.rec_images} is the reconstructed result. The result shows that the reconstructed density is same at the flat density zone whether the blur is considered or not and the reconstructed density is more closer to the true value at the edge between different layers if the blur is considered. Further, the effect range of blur size is reduced to 1-2 pixels. In another word, the blur effect is reduced to an ignorable level.

\subsection{reconstruction of an experimental image}

\begin{figure}[htpb]
\centering
  \includegraphics[height=1.5in]{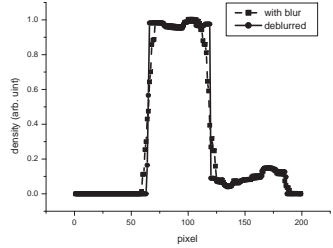}\\
  \caption{The central line of reconstructed density from an experimental image. The solid line with square is result without consideration of blur and the dash line with circle is deblurred result of our forward model.} \label{fig.rec_exp}
\end{figure}

The same method is used to reconstruct density from an experimental image. The constructed result(Fig.\ref{fig.rec_exp}) shows the same result of the reconstruction of the simulated image that when the reasonable blur is considered in our forward model, the effect of blur size is removed completely at the edge between different layers. So the more precise density can be gained with our forward model and the reasonable blur size. The reasonable blur size can be obtained according to our published research.

\subsection{robustness of the method}

\begin{figure}[htpb]
\centering
  \includegraphics[height=1.5in]{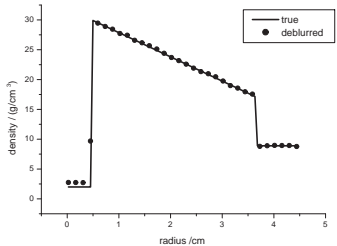}\\
  \caption{The central line of reconstructed density from a simulated image with forward forward. The solid line is true value, the symbol circle is reconstructed result.} \label{fig.rec_sim1}
\end{figure}

Our method was used to reconstruct density from a density-variant object image to investigate the robustness of the method . The density-variant object is formed by changing the unform density of the French Test Object into a sloped one and the density distribution is shown on the Fig.\ref{fig.rec_sim1}. As the process of section\ref{subsec.rec0}, the density distribution is turned into simulated image with FXRMC code and the density is reconstructed from it. The reconstructed result shows that our method is also robust for reconstructing the density-variant object from the simulated image to the given value and can reduce the blur effect to the ignorable level.

\section{Conclusion and discussion} \label{sec.conclusion}
We treated the areal source as many weighted point sources and used the forward model to achieve a reasonable simulated image by comparing to experimental image. Our methods reduced the blur effect on the edge of the density only to 1~2 pixels, and it is also suitable for reconstruction of density-variant object.

Additionally, the determination of the blur size is important to the reconstructed result. If the blur size is much lower than the true value, the effect of blur size on the density can not be removed completely. Otherwise, a faked step will appear in result, detailed in ref\cite{LJIN2015}.
	
The main difference between our model and the LANL model is that we pay more attention on the areal source and handle it as different point sources while LANL pay more attention on the spectrum effect on attenuation coefficient and handle it as many energy bins. In our research, if the object is thick enough that the spectrum effect becomes weak and the attenuation coefficient is almost close to a constant while the blur effect of the source still covers several pixels. So we handle the blur first. In the future, the spectrum effect will be under consideration in the procedure of post-image processing.

{\em Acknowledgements:} This work is supported in part by the Foundation of China Academy of Engineering Physics (No.2013B0202021, No.2014B0403056) and the Youth Natural Sciences Foundation of China (No. 11304295).



%% References


\bibliographystyle{elsarticle-num}
\bibliography{<your-bib-database>}



\end{document}